# An exploratory view of water quality and sediment characteristics in Musa Estuary – the Persian Gulf

S. Mostafa Haghshenas
Oceans Research Co., Tehran, Iran, smostafah1990@yahoo.com

## Abstract

The analysis of physico-chemical and biological characteristics of different water levels and sediment of a water body can make a great exploratory view of environmental conditions; especially in the estuaries. Estuaries are suitable areas to establish industries and harbors for export. Many estuaries facing steady pollution concentration because of distance from the main currents and gyres in the large water bodies. Measurement, comparing and discussion of three important elements including heavy metals, nutrients and benthos in Musa Estuary is the main purpose of this study. Industrial and domestic sewage, petrochemical complexes, and establishment of a large port and it's regarding shipping activities are the main sources of pollution in Musa Estuary. Five sampling points based on the distance from pollution sources was selected. Then three samples from surface water and three samples from bed sediments for each point were taken. In the next step, several physical and chemical tests including flame atomic absorption spectrometry were taken to determine metal concentration. . Highest heavy metal density in surface water (up to 2 ppm) is adjacent to the sources of pollution, petrochemicals and the port.  The level of pollution decreases in the area under severe current speed. Furthermore the density of some elements is higher in sediment samples rather than in water ones, especially for mercury and arsenic. It is due to bed sediment size in this area which is less than 63 micrometer, considered as clay and has a high capacity for accumulation of metals. Two common organic matter pollution metrics indexes; the ratios of Nitrogen to Phosphorus (N/P) and Biological Oxygen Demand to Chemical Oxygen Demand (BOD/COD) have been used. A biological analysis has been conducted on population of benthos and their classification, based on their resistance to the pollution (AMBI factor), collected from mentioned sediment sampling points to investigate the pollution density over the sampling  area.

The results of mentioned physicochemical and biological analysis show that in more polluted sampling points, the number of resistance benthos (Bivalves) are more than other points of the study area. Heavy metal concentration in sediment is between 10 to 1000 times more than water samples especially for Arsenic, Mercury, and Vanadium. Dissolved Oxygen is acceptable all over the study area (more than 4 mg/L) based on environmental quality standards for dissolved oxygen which present the range of 2-9 mg/l for sensitive aquatic lives to less sensitive ones. However the presence of chemicals causes increase in the value of Chemical Oxygen Demand (around 500 to 1000 mg/L) and consequent decrease in water purification ability  Nutrient uptake capacity which shown as the ratio of Nitrogen to phosphorus is appropriate (around 6) in parts of Musa estuary with more than 3 meters depth . The study area is a semi-closed tide-dominant water body with shallow depth, in comparison to the oceans. A 2-D modelling is applied to simulate current speed and direction. The results of this modelling shows except in the creeks, we have relatively suitable circulation of water in the estuary and current speed may exceed 0.8 m/s. It means Musa estuary is not a dead zone and in some parts which have enough water circulation and nutrient uptake for aquatic life, but huge amount of heavy metal (especially Mercury) in the sediment and water samples, indicates that all aquatic  bodies overwhelmed by dangerous heavy metals.

## 1. Introduction

Estuaries are marine embayment that are significantly diluted by fresh water. They are often preferred sites for shipping corridors and harbors construction. Because of shallow depths, slow currents and long water residence time, they are prone to accumulation of pollutants. Estuaries are suitable zones which are used as nursery grounds for commercial aquatic species [1]. Industrial and domestic waste water flows are two main pollution sources that can affect aquatic life; waste water containing heavy metals can have profound effects on marine life. Consumption of metals presents in the body of caught aqua, by the secondary consumers is considered as a serious threat for human health, due to bio-magnification of metals concentration in the food chain.

The heavy metals concentration especially in water bodies could be measured by many different techniques. For gaining a better understanding of these values and the ranges of change in different seas in the world, some studies are mentioned in the following. Many researchers and organizations deal with analysis and monitoring of heavy metals standard levels. Japan International Cooperation Agency (2014) determined acceptable levels of heavy metals in this area [2]. In addition, Jalali *et al*. (2015), Safahieh *et al*. (2013) and Safahieh *et al*. (2011) measured concentrations of some important heavy metals using aquatic organisms as bio indicators in Musa estuary at the northwest corner of the Persian Gulf [3], [4], [5]. In the majority of their results, the values were more than standard levels which were issued by WHO.

Plants, molluscs, and fish are three biological indicators of heavy metalconcentration. The amount of pollution from primary producer (plants) to primary consumer (molluscs), and secondary consumer (fish), exponentially increase [6]. Bivalves are notable species of regions with heavy metal concentration. Their ability to accumulate metals and in some case decrease the level of metal concentration are mentioned in many references. For proof of this ability it would be helpful if there is any correlation between concentration of heavy metal in the body of bivalves and heavy metal concentration in sediment samples from a study area [7].

Estuaries are suitable areas for aquaculture projects. Abundant supply of nutrients, safe distance from severe streams and high wave heights are main good features of estuaries. But an important issue in aquaculture is the health degree of grown species rather than the amount of production. Producing a large amount of fish carrying enormous amount of heavy metals can increase ten times by human consumption. It causes dispersion of toxic materials and cancer outbreak in human life [8].

Bioaccumulation of heavy metals is a critical problem for organisms which lives or grown near petrochemical and industrial complexes. Studies show that the concentration level of metals in soft tissue of bivalves is much higher than sea water and bed sediments. This tremendous amount of metals can grow up in human body after. It is expected the concentration of metal to be increased by going from water to the sediment [9].

Detection of heavy metals have done with several methods of spectrophotometry. Some common approaches including Coupled Plasma – Optical Emission Spectrometry (ICP-OES), flame atomic absorption spectrometer (FAAS), Cold Vapor Atomic Absorption spectrophotometer, furnace atomic absorption spectrophotometry, etc. [6],[10],[11].

In order to investigate the dispersion pattern of pollution in a water body, hydrodynamic features have impressive effects. Residence time and persistency of currents are two important elements which was calculated using Mike 3 model [12]. The analysis of physico-chemical and biological characteristics at different water levels and investigating sediment characteristics can make a great exploratory view of environmental conditions.

In this study, it is tried to collect a comprehensive data set of heavy metal concentrations in different parts of a very important estuary in the Persian Gulf, by taking samples from water and bed sediments, in order to determine effective parameters and evaluate the level of effectiveness on pollution distribution. In addition, taking biological and environmental samples, will help us to find a probable relation between abundance of organisms and physicochemical features of water and sediment.

## 2. Study Area

As a case study, Musa Estuary is a semi closed area with shallow depth (-6 - 0 m), very salty water, and intensively hot and humid weather. Extreme water level changes (around 6 m) and presence of tidal current are two significant features of this water body.

Some potentially sensitive points were considered to take surface water samples and bed sediment samples based on two main criteria: distance to the probable pollution current sources and the effect of extreme current speed. Points 101, 102 and 103 are close to entrance of urban and industrial sewage, point 104 is adjacent to the deepest point of the Persian Gulf with significant current speed, and Control point is considered in a farther location to the pollution sources to monitor changes in concentration values (Figure 1). The samples were taken at the end of summer (October 2017) which maximum level of algal bloom and oxygen consumption are expected due to strong sunlight and increasing photosynthesis. However accumulation of metals and other chemical elements is independent of seasonal changes.

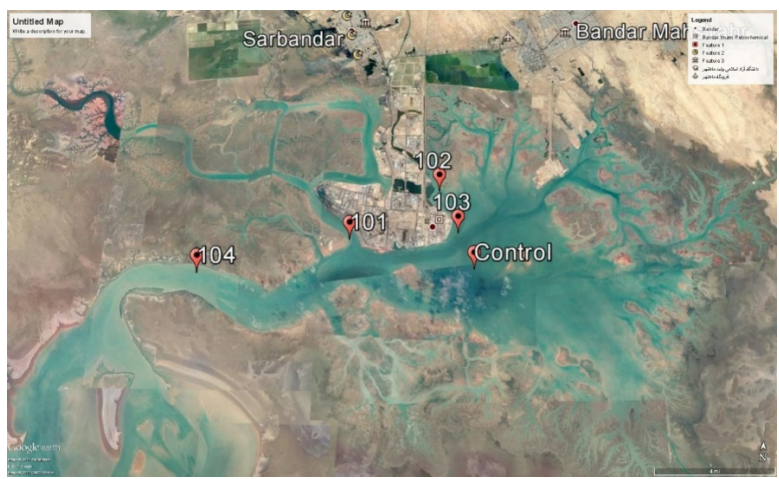

**Figure 1. Distribution of sampling points in the study area**

## 2.1. Heavy Metal

After taking three samples of surface water from each locations, the concentration values of heavy metals wasmeasured by flame atomic absorption device for each sample and average of three values was calculated for every sample points. For sediments, all samples were digested in aqua regia solution, then metal concentrations were determined using mentioned method. In the next step, values were compared with standard levels which mentioned in the report issued by United States Environmental Protection Agency (USEPA) (Table 1 and Table 2) [13]. In water samples, except for Copper, Arsenic, and Mercury, we expose concerning concentrations of metals which can easily affect human body by consuming aquatic organisms and the danger of bio-magnification will increase. In sediment samples, excludinglead and Cadmium, the concentration values for all elements are much more than water samples which illustrates highcapacity of clayparticles to accumulate metals and other pollutants.

**Table 1. Average concentration values and standard levels of heavy metals in water samples**

| Sample ID | Cd (mg/l) | Cr | Co | Cu | Pb | Zn | Ni | As | Hg | Vd |
|---|---|---|---|---|---|---|---|---|---|---|
| Standard | 0.01 | 0.05 | 0.05 | 0.05 | 0.04 | 0.05 | 0.05 | 0.04 | 0.001 | 0.03 |
| 101 | 0.476 | 0.307 | 1.088 | 0.028 | 2.077 | 1.089 | 1.089 | <0.005 | <0.001 | 0.238 |
| 102 | 0.416 | 0.253 | 0.949 | <0.001 | 1.895 | 0.794 | 0.794 | <0.005 | <0.001 | 0.123 |
| 103 | 0.487 | 0.304 | 0.948 | <0.001 | 2.083 | 0.83 | 0.83 | <0.005 | <0.001 | 0.234 |
| 104 | 0.472 | 0.304 | 0.966 | <0.001 | 2.057 | 1.011 | 1.011 | <0.005 | <0.001 | 0.289 |
| Control | 0.497 | 0.312 | 0.969 | 0.007 | 2.223 | 0.994 | 0.994 | <0.005 | <0.001 | 0.087 |

**Table 2. Average concentration values and standard levels of heavy metals in sediment samples**

| Sample ID | Cd (mg/kg) | Cr | Co | Cu | Pb | Zn | Ni | As | Hg | Vd |
|---|---|---|---|---|---|---|---|---|---|---|
| Standard | 1.5 | 43 | 2 | 31 | 50 | 121 | 21 | 0.025 | 0.15 | 57 |
| 101 | <0.15 | 36.56 | 10.12 | 5.52 | <1 | 107.04 | 56.32 | 4.18 | 10.6 | 81.3 |
| 102 | <0.15 | 37.28 | 10.4 | 5.28 | <1 | 88.44 | 56.48 | 4.33 | 10.5 | 79.6 |
| 103 | <0.15 | 31.44 | 8.48 | 2.6 | <1 | 90.44 | 48.28 | 4.73 | 10.7 | 72.3 |
| 104 | <0.15 | 73.76 | 27.12 | 16.44 | <1 | 115.32 | 132.04 | 4.83 | 9.89 | 76.7 |
| Control | <0.15 | 65.44 | 24.8 | 11.92 | <1 | 106.12 | 98.88 | 5.48 | 3.89 | 67.5 |

To have a better visual understanding of pollution dispersion, concentration data and the location of sampling points were defined as information layers in Arc GIS. Then by data interpolation and making a boundary around sampling area the changes in heavy metal levels are illustrated in Figure 2-a to Figure 2-f for each element of heavy metals in surface water samples. Accumulation near to the sources of pollution is expected, but there is a weird increase in concentration of metal in control point which may occur because of current pattern.

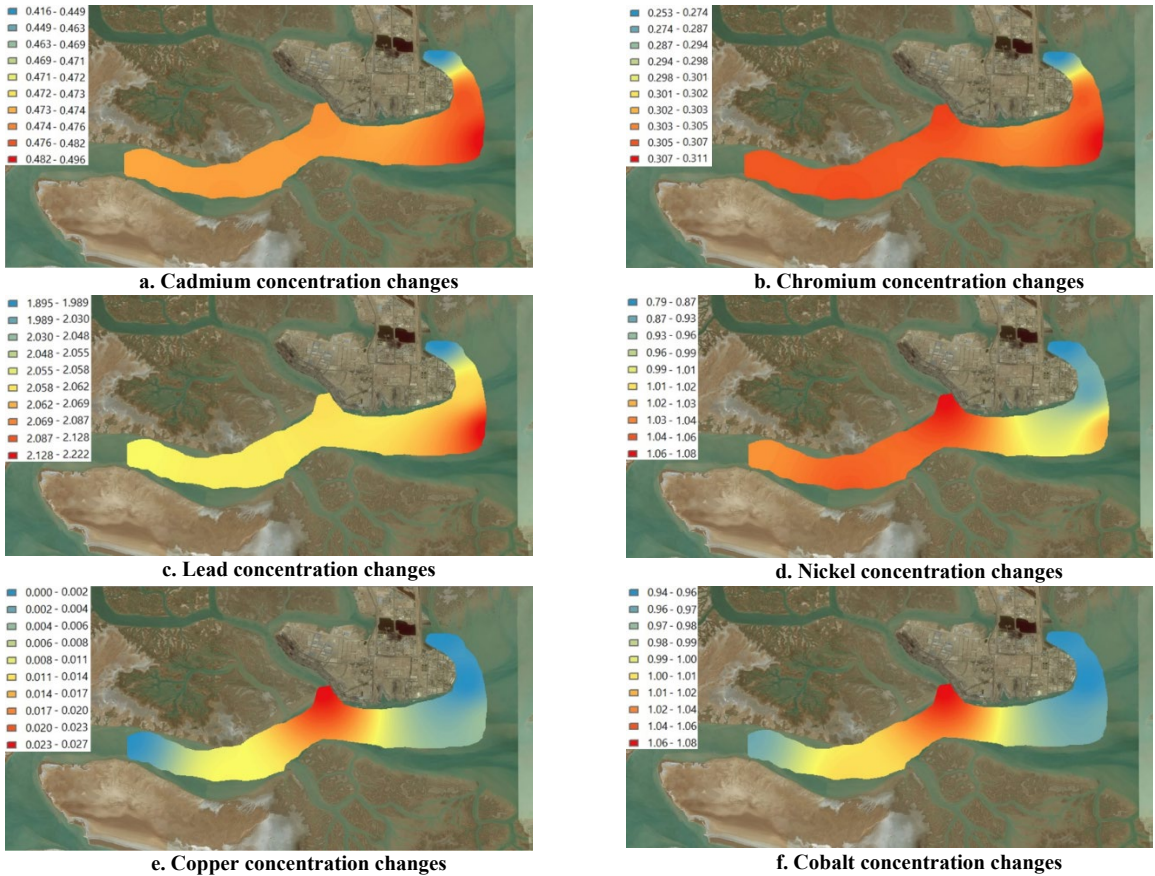
**Figure 2. Changes in heavy metal levels of heavy metals in surface water**

## 2.2. Physicochemical Characteristics

The summary of physicochemical characteristics of surface water samples in the study area is provided in Table 3. Due to enormous importance of dissolved oxygen in aquatic health, it is vital to provide an appropriate insight into oxygen consumption reactions. The sampling time was end of summer which maximum level of algal bloom and oxygen consumption are expected due to strong sunlight and increasing photosynthesis. However, the oxygen levels seem appropriate in this study in comparison with international standards [14]. For biological oxygen demand the condition is same; except in point number 104 where maybe the existence of organisms for oxygen consumption is less. For nitrogen and phosphorus, results show high concentration values. Also chemical oxygen demand values were measured at concerning levels; more than around 20 times than standard level. Based on EPA standards, suitable range of pH for most of aquatic organisms is between 6.5 to 8 ppm. In this study area this parameter has the value up to 8.5 ppm which can reduce the diversity of organisms and reproduction ability of water. For aquaculture, this range can reach 9 [15].

**Table 3. Physicochemical characteristics of water samples in the study area**

| Points | Nitrogen | Phosphor | pH | Sulfate | Temp. | COD | BOD | Cond. | Sal. | Turb. | DO |
|---|---|---|---|---|---|---|---|---|---|---|---|
| Unit | ppm | ppm | ppm | ppm | °C | mg/L | mg/L | ms/cm | ppt | NTU | mg/L |
| Standard | 0.4 | 0.045 | | 2712 | | 50 | 20 | | | | 4 |
| 101 | 0.9 | 0.27 | 8.3 | 4700 | 30 | 1080 | 4.22 | 73.9 | 44.0 | 28 | 4.69 |
| 102 | 3 | 0.51 | 8.1 | 4100 | 30 | 570 | 8.43 | 64.4 | 42.4 | 38 | 5.32 |
| 103 | 1.3 | 0.16 | 8.5 | 4200 | 29 | 1320 | 18.97 | 75.1 | 44.2 | 39 | 5.32 |
| 104 | 0.9 | 0.1 | 8.5 | 4050 | 30 | 640 | 39.99 | 73.5 | 44.0 | 60 | 5.86 |
| Control | 1 | 0.16 | 8.4 | 4150 | 29 | 1160 | 12.67 | 75 | 44.2 | 34 | 4.15 |

## 2.3. Benthos

Biological analysis based on bio-indicators is a common method researches. Samples are taken from bed sediments in the mentioned locations and then are stabilized with commercial formaldehyde. Next step of sample preparation is sieving with 18 mesh to get rid of huge amount of very small particles of clay (around 90 percent of each sample). In Figure 8 to Figure 14 organisms are illustrated under 20x Loupe Magnifier.

The relation between abundance of organisms is used to categorize them in biological groups according to a continuous index [16], [17]. The AMBI factor represents a formula to show the tolerance of biological groups toward severity of pollution in the study area.

$$AMBI = (0* GI) + (1.5*GII) + (3*GIII) + (4.5*GIV) + (6*GV) / 100$$

Based on the mentioned biotic index, Group I: Including the most sensitive species to the pollution which represents unpolluted areas, Carnivores and some deposit-feeder's polychaets. Group II: Including less sensitive species and represents low densities of pollutants with low changes with time, suspension feeders, less selective carnivores and scavengers are considered in this group. Group III: Including tolerant species to higher amount of organic matter, these species may endure under normal condition; suspension deposit-feeders are categorized in this group. Group IV: Second order opportunistic species, mainly small polychaets, subsurface deposit feeders. Group V: First order opportunistic species, mainly deposit-feeders. In, the organisms have been sorted according to the mentioned ecological groups from Borja et al., (2000) [18].

For the value of AMBI factor between 3.3 to 4.3, the study area is considered as Meanly Polluted which we can see for point numbers 102, 103, and the control point. Two points number 101 and 104 are unpolluted; the reason will be obtained in discussion.

**Table 4. Detected organisms in bed sediments**

| AMBI | 0 | | AMBI | 3.3 | | AMBI | 3.75 | | AMBI | 0 | | AMBI | 3.9 | |
|---|---|---|---|---|---|---|---|---|---|---|---|---|---|---|
| 101 | Group | Abundance | 102 | Group | Abundance | 103 | Group | Abundance | 104 | Group | Abundance | Control | Group | Abundance |
| Isopods | I | I | Gastropods (*Fissurella*) | I | I | Bivalves (*Ceratocardium*) | III | II | Gastropods | I | I | Bivalves | V | III |
| | | | Bivalves | V | II | Polychaetes | IV | II | Detritus | | I | Gastropods | I | I |
| | | | Hydrozoas | IV | I | | | | | | | Polychaetes | IV | IV |
| | | | Gastropods | I | I | | | | | | | Sea pen (Coral) (*Virgularia*) | I | I |
| | | | | | | | | | | | | Polychaete (*Sternaspis scutata*) | III | I |

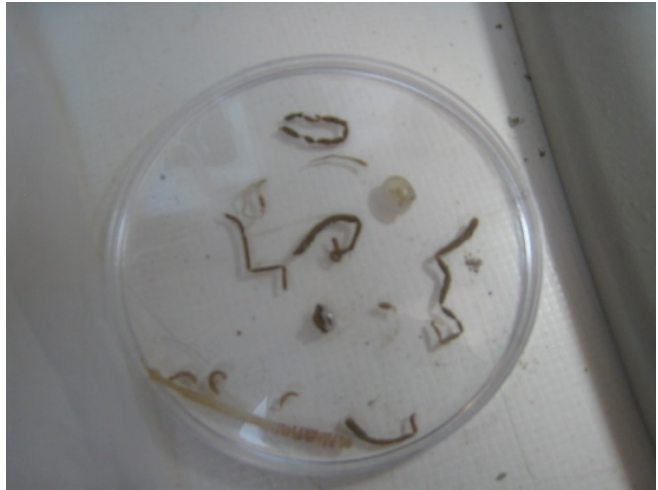
-a- All detected organisms

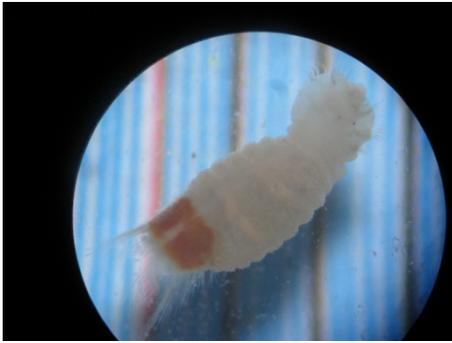
b- Sternaspis Scutata

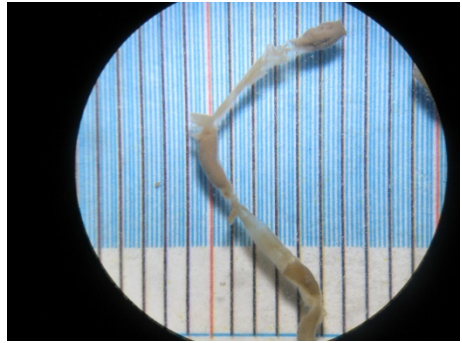
c- Polychaete in a tunnel made by itself

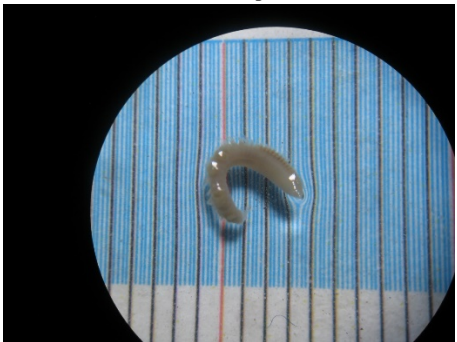
d- Polychaete

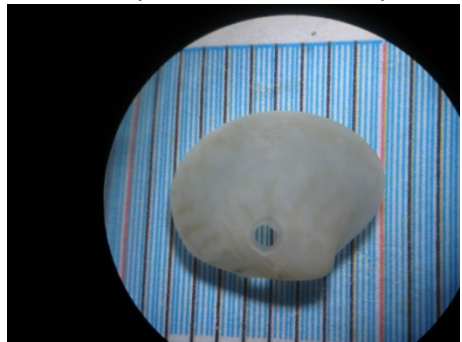
e- Fissurella (Dead Orgnism)

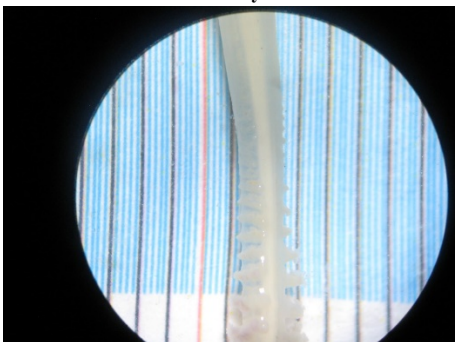
f- Virgularia

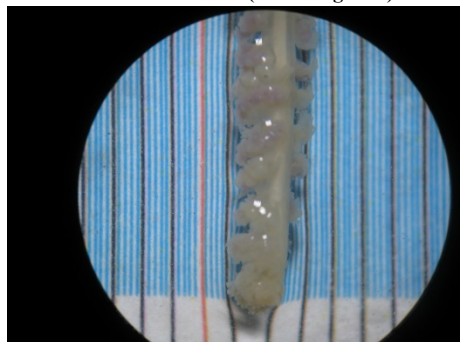
g- Virgularia

**Figure 3. The detected benthos in the study area**

## 3. Discussion

The interaction between physicochemical characteristics and heavy metal concentrations is a key parameter in seawater and bed sediment analysis [19]. In following study, Statistica V12 software was run (P<0.05). The values more than +0.5 and less than -0.5 are considered as significant values, with direct and reverse relations respectively (Table 5). Then another analysis on correlations between heavy metal concentrations and abundance of organisms in sediment samples are applied using the mentioned software (Table 6). Organisms could be used as bio-indicator of pollutants, so that their tolerance to pollution differ from each other and the level of contamination of each area could be determined [20].

**Table 5. Correlation matrix between all water parameters analyzed in the study area**

|     | T | pH | S | DO | COD | BOD | TP | TN | Cd | Cr | Co | Cu | Pb | Zn | Ni | Vd |
|---|---|---|---|---|---|---|---|---|---|---|---|---|---|---|---|---|
| T   | 1.000000 | | | | | | | | | | | | | | | |
| pH  | -0.4910 | 1.000000 | | | | | | | | | | | | | | |
| S   | -0.5238 | 0.8807 | 1.000000 | | | | | | | | | | | | | |
| DO  | 0.4609 | 0.1509 | -0.2652 | 1.000000 | | | | | | | | | | | | |
| COD | -0.7887 | 0.4865 | 0.7178 | -0.5611 | 1.000000 | | | | | | | | | | | |
| BOD | 0.0674 | 0.6724 | 0.3036 | 0.6847 | -0.3072 | 1.000000 | | | | | | | | | | |
| TP  | 0.4482 | -0.9719 | -0.9284 | -0.0021 | -0.4714 | -0.6261 | 1.000000 | | | | | | | | | |
| TN  | 0.2744 | -0.8083 | -0.9567 | 0.2271 | -0.5375 | -0.3424 | 0.9001 | 1.000000 | | | | | | | | |
| Cd  | -0.6488 | 0.8258 | 0.9798 | -0.4185 | 0.7823 | 0.1969 | -0.8860 | -0.9083 | 1.000000 | | | | | | | |
| Cr  | -0.4516 | 0.8253 | 0.9891 | -0.3333 | 0.6728 | 0.2631 | -0.9057 | -0.9808 | 0.9718 | 1.000000 | | | | | | |
| Co  | 0.3951 | -0.1344 | 0.2534 | -0.3712 | 0.2042 | -0.4381 | 0.0070 | -0.4223 | 0.1967 | 0.3458 | 1.000000 | | | | | |
| Cu  | 0.2669 | -0.1683 | 0.2661 | -0.5258 | 0.3081 | -0.5573 | 0.0317 | -0.4051 | 0.2437 | 0.3593 | 0.9837 | 1.000000 | | | | |
| Pb  | -0.6733 | 0.6561 | 0.8657 | -0.6314 | 0.7024 | 0.0722 | -0.7698 | -0.8096 | 0.9397 | 0.8883 | 0.1574 | 0.2413 | 1.000000 | | | |
| Zn  | -0.5198 | -0.1793 | -0.3856 | 0.0817 | 0.1252 | -0.1618 | 0.3439 | 0.6342 | -0.2871 | -0.4931 | -0.7683 | -0.6886 | -0.2667 | 1.000000 | | |
| Ni  | 0.2289 | 0.3050 | 0.5866 | -0.3435 | 0.1403 | 0.0751 | -0.4920 | -0.7831 | 0.5361 | 0.6926 | 0.7569 | 0.7263 | 0.5540 | -0.9454 | 1.000000 | |
| Vd  | 0.3610 | 0.5337 | 0.3842 | 0.6628 | -0.0392 | 0.5795 | -0.4643 | -0.4713 | 0.1935 | 0.3513 | 0.2746 | 0.1231 | -0.1094 | -0.4721 | 0.3140 | 1.000000 |

There are some statistically significant correlations between physical and chemical characteristics and metals concentration values which can be useful to categorize heavy metals. One category is Cobalt, and Copper with same r values with Zinc and Nickel. Another category is Cadmium, Chromium, and Lead which have same r values with salinity, pH, total Nitrogen and Phosphorus. This correlation illustrate the same source of metals for each category [19].

**Table 6. Correlation matrix between Heavy Metal concentration values in sediment and abundance of micro-organisms based on biotic index in the study area**

|     | Cr | Co | Cu | Zn | Ni | As | Hg | Vd | Grp I | Grp III | Grp IV | Grp V |
|---|---|---|---|---|---|---|---|---|---|---|---|---|
| Cr      | 1.000000 | | | | | | | | | | | |
| Co      | 0.997419 | 1.000000 | | | | | | | | | | |
| Cu      | 0.986697 | 0.972742 | 1.000000 | | | | | | | | | |
| Zn      | 0.781058 | 0.764064 | 0.823675 | 1.000000 | | | | | | | | |
| Ni      | 0.983807 | 0.970987 | 0.992900 | 0.800679 | 1.000000 | | | | | | | |
| As      | 0.672433 | 0.722423 | 0.543913 | 0.298411 | 0.569631 | 1.000000 | | | | | | |
| Hg      | -0.570955 | -0.622313 | -0.449272 | -0.300045 | -0.417203 | -0.872578 | 1.000000 | | | | | |
| Vd      | -0.383336 | -0.445801 | -0.231180 | -0.024620 | -0.273658 | -0.938458 | 0.787109 | 1.000000 | | | | |
| Grp I   | 0.347280 | 0.350123 | 0.329148 | 0.068358 | 0.238328 | 0.217557 | -0.551771 | -0.036185 | 1.000000 | | | |
| Grp III | -0.266568 | -0.211128 | -0.391071 | -0.420559 | -0.309062 | 0.445512 | -0.195762 | -0.713787 | -0.534522 | 1.000000 | | |
| Grp IV  | 0.152416 | 0.220521 | -0.008007 | -0.213026 | -0.000997 | 0.805548 | -0.831419 | -0.908873 | 0.285714 | 0.634745 | 1.000000 | |
| Grp V   | 0.242315 | 0.280741 | 0.142118 | -0.185102 | 0.086525 | 0.539183 | -0.776675 | -0.494253 | 0.845154 | 0.000000 | 0.739510 | 1.000000 |

There are some statistically significant correlations between metal concentration values and abundance of each group. Especially for Arsenic, Mercury, and Vanadium. Mercury and Vanadium have reverse significant relation with three groups of organisms. But Arsenic has direct significant relation with groups IV and V of organisms. Arsenic also has reverse significant relation with Mercury and Vanadium.

The organisms in groups III, IV, and V including polychaetes and bivalves uses metals as antipredation [21]. Mentioned organisms are able to produce chemical compounds of metals against different threats. These organisms specially Bivalves are considered as resistant species to hyper-accumulation of metals [22].

Vanadium is one of the elements which shows the highest concentrations rather than other metals in the Persian Gulf [22]. Wastewater from oil tankers, oil leakage from drilling rigs and some other environmental events are the main sources of pollution.

Bivalves (organisms in groups III and V) could be beneficious in water purification. Their ability to eliminate Mercury is different between one species to another [23]. The capacity of Mercury accumulation and elimination by Bivalves also is affected by some factors including uptake rate of Mercury by bivalves, initial concentration of accumulated mercury, etc. In the study of Langston, W. J. on 1982, straight relations was observed between amount of organic contents and Mercury concentrations in digested samples of a bivalve species body. Accumulation of Mercury in the body of organisms is directly related to pollutions coming from human activities [24].

Some species have the ability of heavy metal elimination. Especially in bivalves, impressive effect on mercury removal in a deposit-feeding bivalve was measured by Riisgard et. al., 1985. By transferring bivalves from clean water to abroad of an industrial area, Mercury concentration has been rapidly increased in the body of this organism but Mercury absorption gradually decrease after 30 days of exposure [25]. Methylation of heavy metals consider as a treating reaction of pollutants. For instance a kind of mediterranean polychaete able to do the mentioned reaction on Arsenic and transform it to dimethylarsenic acid, a moderately toxic compound [26].

### 3.1. Modelling

We also develop a 2D hydrodynamic model to make a visual analysis of these effective parameters. To set up a 2D hydrodynamic model in Mike- DHI software, different parameters including: bathymetry data and initial and boundary conditions were considered; then the unstructured mesh was applied. For model calibration various sensitivity tests were done to examine roughness heights and eddy viscosity coefficient; then measured and simulated tidal elevation was compared. Finally, daily changes of current direction and speed with a 15-minute time step is shown in the model output (Figure 4).

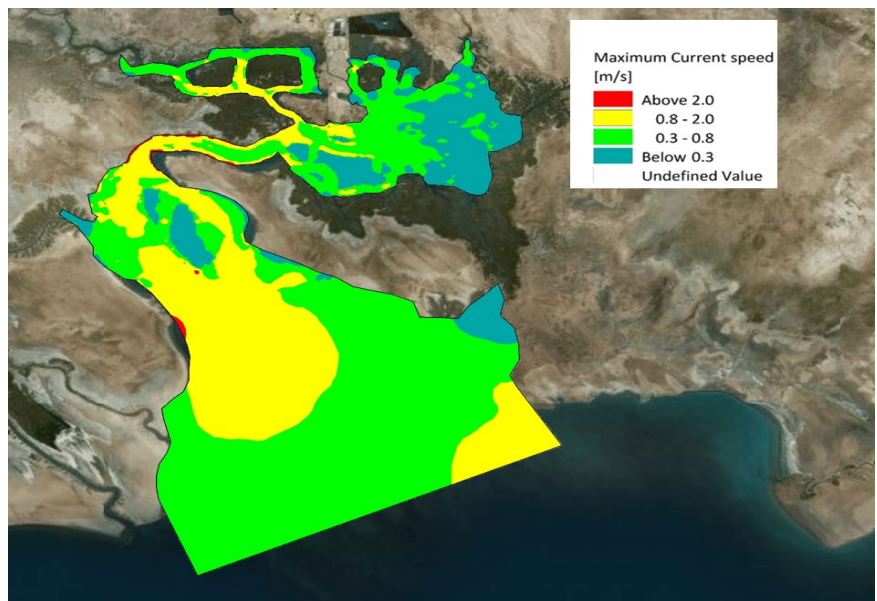

**Figure 3. Results of current speed modelling in Musa estuary**

For Control point, it seems a contrast between distance to the pollution source and heavy metal accumulation; although this point is farther, values of chromium, cadmium, and lead are much more compared with other points in the study area which refers to mentioned decrease in the current speed. Control point is the start of decrease in current speed and depth by moving to the east. So the indicated increase in accumulation of metal in control point and other shallow waters is expected.

Current direction is another important effect in pollution accumulation. It changes approximately every six hours (Figure 5). Tidal current and water flux in entrance of the estuary are two main parameters which affect current direction.

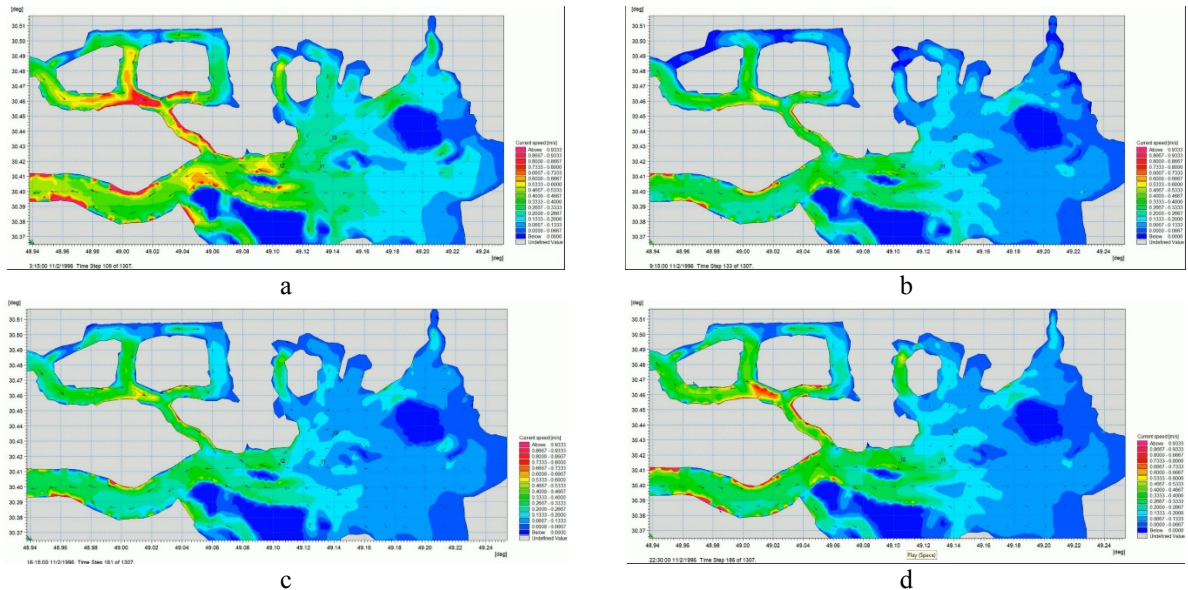

**Figure 4. Change in water direction during a day in Musa Estuary**

### 3.2. Important Ratios

In that study, some scenarios defined for eutrophic zones where remarkable presence of nutrients with high consumption of oxygen occur in different parts of the study area. Furthermore, Z.Wan et al. (2011), and Hee Jeong Choi et al. (2014) used a criteria for the balance of mass productivity and nutrient removal as the ratio of nitrogen (N) to phosphorus (P) [27],[28]. According to their results, the values between 5-30 for ratio of Nitrogen to Phosphorus are acceptable for sea water. In addition, Curtis Deutsch et al. (2012) introduced a standard level for N to P ratio around 15 for the oceans. For biological purification ability of water, Khaled Zaher et al. (2015) used the ratio of BOD to COD [29], [30]. For values more than 0.6, water can purified biologically, for the values between 0.3 to 0.6 the treatment process rate is slow, and if the ratio is less than 0.3 biological purification do not proceed due to having some toxic components and non-biodegradable substances or water acclimated microorganisms may be required in its stabilization.

**Table 7. Parameters of water quality in the study area**

| Sample ID | Total Nitrogen (ppm) | Total Phosphorus (ppm) | N/P | BOD (mg/l) | COD (mg/l) | BOD/COD |
|---|---|---|---|---|---|---|
| Standard | 0.4 | 0.045 | 15 | 20 | 5 | 0.3< |
| 101 | 0.9 | 0.27 | 2.36 | 4.22 | 1080 | 0.003 |
| 102 | 3 | 0.51 | 4.16 | 8.43 | 570 | 0.014 |
| 103 | 1.3 | 0.16 | 5.75 | 18.97 | 1320 | 0.014 |
| 104 | 0.9 | 0.1 | 6.37 | 39.99 | 640 | 0.062 |
| Control | 1 | 0.16 | 4.42 | 12.67 | 1160 | 0.01 |

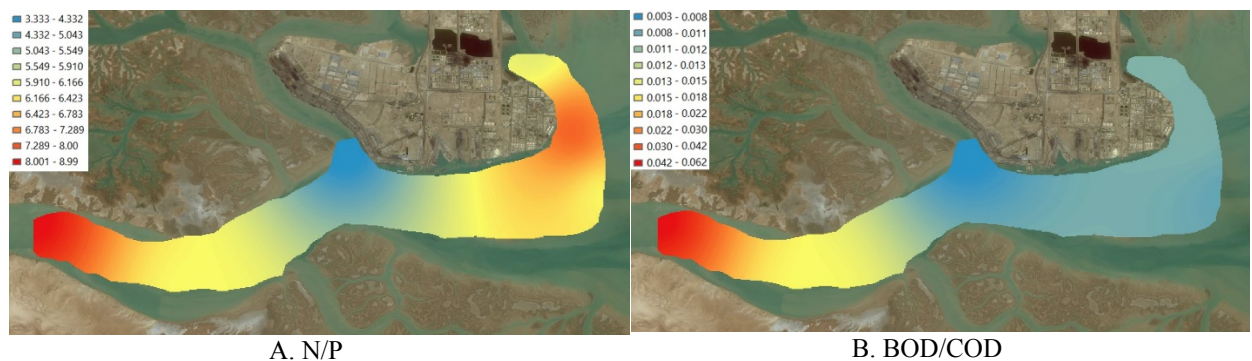
A. N/P                  B. BOD/COD

**Figure 5. Change in N/P and BOD/COD in study area**

### 3.3. Heavy Metal Dispersion

Changes of concentration values for heavy metals represented by two distinct trends. For Cadmium, Chromium, and Lead the pollutants are accumulated around the control point (Figures 2 to 4). For Copper, Cobalt, and Nickel the most concentration values is obvious around the point number 101, which is one of the sources of pollutant current entrance (Figures 5 to 7). For the first trend, the main reason of accumulation is related to decrease in current speed. For the second one, metals accumulation is shown around the point number 101 (Figure 1), and by moving to the north, the current speed will increase and water current carries metal contents into the creeks. Then in continue to the end of the creeks the current will slow down and metals accumulation may extend.

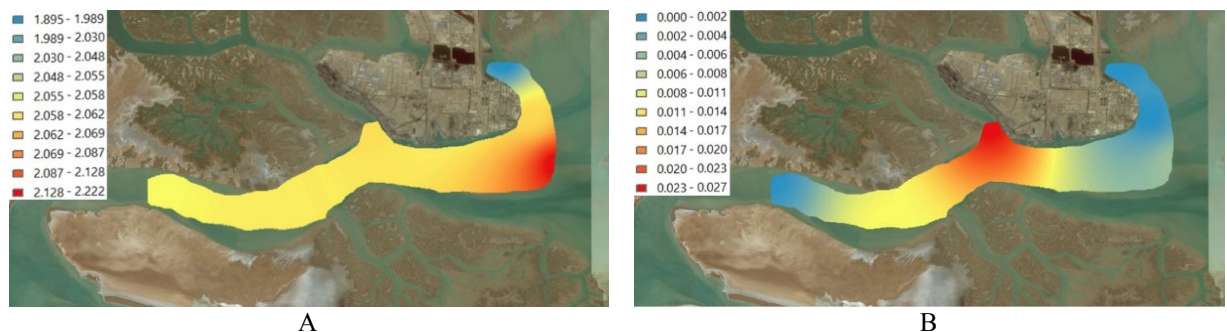
A                  B

**Figure 6. Two main trends for changes in heavy metal concentration in surface sea water in the study area**

## 4. Conclusion

Water quality and sediment characteristics are very essential for any environmental study [31], [32]. These data are necessary for Environmental Impact Assessment [33], aquaculture [34], and any other development in the marine and coastal areas. Sediment origin is also an important factor for obtaining a comprehensive understanding of the study area [35]. Moreover, physical oceanography data of winds, waves, and currents are necessary to be available in the Persian Gulf for defining the boundaries of Musa Estuary, which have been studied by different researchers [36], [37], [38]. Here, describing initial environmental features of the study area, it is tried to demonstrate the relations between pattern of change in physicochemical parameters, statistical analysis and current modelling. In addition, the abundance of benthos species may have a close relation with concentration values of metals.

Due to huge amount of phosphorus in the study area, it was expected to see a massive eutrophication. However the level of dissolved Oxygen (around 5 mg/l) and pH (around 8 ppm) show that this area is not a dead zone and aquatic life could be supported in Musa Estuary. However high level of metals threat secondary users is a food chain. About the water quality of the study area, COD is in the most critical situation. In three sampling points this parameter measured more than 1000 ppm which we expect this amount in sewage. For the bed sediment, extremely concerning amount of Mercury, Arsenic, and Vanadium, reject any kind of fish farming in such a toxic environment. The concentration values of these three elements is some case are around 100 times more than the standard levels.

Based on GIS figures, for surface water samples, three elements have same approaches of dispersion. These elements including Cadmium, Chromium, and Lead, have same r values in correlation table and also same approach of dispersion in the study area. The values of Cadmium, Chromium, and Lead concentrations, show high amounts of positive r values (+0.78, +0.67, +0.7) respectively with the values of Chemical Oxygen Demand. This indicates high amounts of COD, affected by high concentration values of these three elements. For sediment samples, three elements including Mercury, Vanadium, and Mercury have significant relationship with benthos groups. Especially the groups number IV and V which are more resistant to the pollution. The r value between Mercury and Vanadium concentration values and IV and V groups of benthos varies between -0.77 and -0.9. However, the r values between Arsenic concentration values and IV and V groups of benthos shows positive amounts (+0.54 to +0.8). This difference in biological reaction of these three mentioned elements, comes from different nature of them. Each element produced by a specific petrochemical reaction and definitely have specific solution to reduce and control.

Current modelling of this water body demonstrated the border of area which has suitable water circulation. Furthermore, some important points where accumulation increase and change in water current direction occurred, are determined. From the control point (Figure 1) toward the creeks, the current speed decrease; consequently good water circulation is not expected. So, it could be concluded that from the control point to the entrance of the estuary, suitable water current speed range (between 0.3 m/s to 0.7 m/s) is illustrated (Figure 5).

Obviously, reducing production of pollutants from the industrial and domestic sources is the most effective way to decrease toxicity of a water body. However some organisms can play an important role in water purification. In future studies, it worth to investigate methods of metal elimination using biological treatments. Harvesting some of aquatic organisms, especially bivalves, may help to reach this aim. It is highly recommended to investigate which organisms are resistant to the heavy metals. They can survive in toxic areas and also are able to accumulate metals in their body and reduce the toxicity of water by some reactions. Arsenic, Mercury, and Vanadium, as three critical elements in heavy metal group of pollutants, should be analyzed separately. First of all their dispersion range through the Persian Gulf based on specific initial and boundary conditions. After that, the solutions of reducing their concentration and controlling their dispersion. Finally methods to reduce intake in the Persian Gulf such as treatment from the source of pollutions, petrochemicals and refineries, and some approaches of reducing concentration values of metals by biological reactions should be investigated. For Musa Estuary, it would be helpful to gather comprehensive historical environmental and geophysical data and comparing with this article to make a better view in potential of development in this area.

## 5. References

[1] T. Cohen, S. S. Q. Hee, and R. F. Ambrose, "Trace metals in fish and invertebrates of three California Coastal Wetlands," Mar. Pollut. Bull., vol. 42, no. 3, pp. 224–232, 2001.
[2] J. I. C. Agency, "The Project for Strengthening Environmental Management in Petroleum Industry in Persian Gulf and Its Coastal Area," Japan Int. Coop. Agency, 2014.
[3] K. S. K Jalali, B Abtahi, "Assessment of cd level in liver and muscle tissues of platycephalus indicus and sediment of musa estuary (northwest of persian gulf)," 2015.
[4] A. Safahieh, "Assessment of Mercury Intake through Consumption of Yellowfin Seabream (Acanthopagrus latus) from Musa Estuary," vol. 1, no. 2, pp. 142–146, 2013.
[5] F. A. Monikh, M. T. Ronagh, A. Savari, and A. S. Area, "Determination of heavy metals (Cd , Co , Cu , Ni and Pb) in croacker fish (Johnius belangerii) from Musa estuary in the Persian Gulf," vol. 2, no. 6, pp. 2–7, 2011.
[6] I. G. David, M. L. Matache, A. Tudorache, G. Chisamera, L. Rozylowicz, and G. L. Radu, "Food chain biomagnification of heavy metals in samples from the lower prut floodplain natural park," Environ. Eng. Manag. J., vol. 11, no. 1, pp. 69–73, 2012.
[7] K. L. S. Cheraghi M., Espergham O., Nooriaee M H., "Assessment of oyster Crassostrea gigas as Biomonitor Agent for Some Metals (Pb and Cu) from Musa Estuary (Persian Gulf)," vol. 793, no. 2, pp. 55–60, 2013.
[8] F. Abdolahpur Monikh, A. Safahieh, A. Savari, and A. Doraghi, "Heavy metal concentration in sediment, benthic, benthopelagic, and pelagic fish species from Musa Estuary (Persian Gulf)," Environ. Monit. Assess., vol. 185, no. 1, pp. 215–222, 2013.
[9] A. Begum, S. Harikrishna, and I. Khan, "Analysis of Heavy metals in Water , Sediments and Fish samples of Madivala Lakes of Bangalore , Karnataka .," vol. 1, no. 2, pp. 245–249, 2009.
[10] F. A. Monikh et al., "Heavy Metals Levels in Sediment and Ray Fish (Dasyatis bennettii) from Musa Estuary and Selech Estuary, Persian Gulf," Am. J. Toxicol. Sci., vol. 3, no. 4, pp. 224–227, 2011.
[11] N. Jaafarzadeh Haghighi Fard et al., "Determination of nickel and thallium concentration in Cynoglossus arel

fish in Musa estuary, Persian Gulf, Iran," Environ. Sci. Pollut. Res., vol. 24, no. 3, pp. 2936–2945, 2017.
[12]     A. Payandeh, N. H. Zaker, and M. H. Niksokhan, "Numerical modeling of pollutant load accumulation in the Musa estuary, Persian Gulf," Environ. Earth Sci., vol. 73, no. 1, pp. 185–196, 2014.
[13]     National Recommended Water Quality Criteria - Aquatic Life Criteria Table. United States Environmental Protection Agency.
[14]     N. G. C. and R. I. C. M J Stiff, "Environmental Quality Standards for Dissolved Oxygen," 1992.
[15]     W. A. Wurts and R. M. Durborow, "Interactions of pH , Carbon Dioxide , Alkalinity and Hardness in Fish Ponds," vol. 0, no. 464, pp. 1–4, 1992.
[16]     A. Borja, J. Franco, and M. Environment, "A Marine Biotic Index to Establish the Ecological Quality of Soft-Bottom Benthos Within European Estuarine and Coastal Environments," vol. 40, no. 12, 2000.
[17]     J. Grall and M. Glemarec, "Using biotic indices to estimate macrobenthic community perturbations in the Bay of Brest," Estuar. Coast. Shelf Sci., vol. 44, no. SUPPL. A, pp. 43–53, 1997.
[18]     S. Dehghan Madiseh, F. Esmaily, J. G. Marammazi, E. Koochaknejad, and S. Farokhimoghadam, "Benthic invertebrate community in Khur-e-Mussa creeks in northwest of Persian Gulf and the application of the AMBI (AZTI's Marine Biotic Index)," Iran. J. Fish. Sci., vol. 11, no. 3, pp. 460–474, 2012.
[19]     H. M. Faragallah, A. I. Askar, M. A. Okbah, and H. M. Moustafa, "Physico-chemical characteristics of the open Mediterranean sea water far about 60 Km from Damietta harbor, Egypt," J. Ecol. Nat. Environ., vol. 1, no. 5, pp. 106–119, 2009.
[20]     "Indicators: Benthic Macroinvertebrates," United States Environ. Prot. Agency.
[21]     M. N. Daniele Fattorini, Alessandra Notti and F. Regoli, "Hyperaccumulation of vanadium in the Antarctic polychaetePerkinsiana littoralisas a natural chemical defenseagainst predation.pdf." .
[22]     S. M. Moosavian, S. M. Baghernabavi, E. Zallaghi, M. Mohammadi Rouzbahani, E. Hosseini Panah, and M. Dashtestani, "Measurement of Pollution Levels Caused by Heavy Metals of Vanadium, Nickel, Lead and Copper Using Bivalve Shells of Timoclea imbricata Species On Bahrakan Coast in Spring 2013," Jundishapur J. Heal. Sci., vol. 6, no. 4, pp. 0–4, 2014.
[23]     H. U. Riisgard and F. Mehlenberg, "Accumulation, elimination and chemical speciation of mercury in the bivalves," vol. 62, pp. 55–62, 1985.
[24]     W. J. Langston, "THE DISTRIBUTION OF MERCURYIN BRITISH ESTUARINE SEDIMENTS AND ITS AVAILABILITY TODEPOSIT-FEEDING BIVALVES.pdf." 1982.
[25]     H. U. Riisgard and F. Mehlenberg, "Accumulation, eliinination and chemical speciation of mercury in the bivalves," vol. 62, pp. 55–62, 1985.
[26]     F. R. Alessandra Notti, Daniele Fattorini, Erika M Razzetti, "BIOACCUMULATION AND BIOTRANSFORMATION OF ARSENIC IN THEMEDITERRANEAN POLYCHAETESABELLA SPALLANZANIIEXPERIMENTAL OBSERVATIONS." 2006.
[27]     H. J. Choi and S. M. Lee, "Effect of the N/P ratio on biomass productivity and nutrient removal from municipal wastewater," Bioprocess Biosyst. Eng., vol. 38, no. 4, pp. 761–766, 2015.
[28]     Z. Wan, L. Jonasson, and H. Bi, "N/P ratio of nutrient uptake in the Baltic Sea," Ocean Sci., vol. 7, no. 5, pp. 693–704, 2011.
[29]     C. Deutsch and T. Weber, "Nutrient Ratios as a Tracer and Driver of Ocean Biogeochemistry," Ann. Rev. Mar. Sci., vol. 4, no. 1, pp. 113–141, 2011.
[30]     Abdalla and Hammam, "Correlation between Biochemical Oxygen Demand and Chemical Oxygen Demand for Various Wastewater Treatment Plants in Egypt to Obtain the Biodegradability Indices," Int. J. Sci. Basic Appl. Res., vol. 13, no. 1, pp. 42–48, 2014.
[31]     A. Razavi Arab, Danehkar, A., Monavari, S. M., Haghshenas S. A. (2013): Monitoring Coral Ecosystems in Naiband Bay, the Persian Gulf, In: Caspian Journal of Applied Sciences Research, Vol. 2, No. 9, pp. 7 - 9, ISSN 2251-9114.
[32]     A. Razavi Arab, Danehkar A., Haghshenas S. A. (2012): Assessment of Coastal Development Impacts on Coral Ecosystems in Naiband Bay, the Persian Gulf, 33rd Coastal Eng. Conf., ASCE, Santander, Spain.
[33]     M. J. Risk (Invited Keynote Speaker), Haghshenas S. A., Razavi Arab Azadeh, (2014): Biology meets engineering on the Iranian Persian Gulf coastline, 11th International Conference on Coasts, Ports & Marine Structures, ICOPMAS, Tehran, Iran.
[34]     S. A. Haghshenas, Haghighi. S., Razavi Arab, A., Shojaee Borhan, S. M., Risk, M. J., Bakhtiari, A., Jedari Attari, M. (2015): Feasibility Study for Developing an Environmentally Sustainable Integrated Multi-Trophic Mariculture System in the Northern Persian Gulf and Gulf of Oman Coastlines, Book of abstracts, Middle East Aquaculture Forum – MEAF 2015, Dubai, UAE.
[35]     A. Razavi Arab, Haghshenas S. A., Samsami, F., Risk, Michael John (2015): Traces of sediment origin in

rheological behavior of mud samples taken from the North-Western Persian Gulf, 13th International Conference on Cohesive Sediment Transport Processes, INTERCOH 2015, Leuven, Belgium.
[36]   A. Razavi Arab, Haghshenas, S. A., Zaker, H. (2015): Deep water current velocity data in the Persian Gulf, Book of abstracts, European Geosciences Union General Assembly, EGU2015, Vienna, Austria.
[37]   M. Jedari Attari, Haghshenas S. A., Soltanpour, M., Allahyar, M. R., Ghader, S., Yazgi, D., Razavi Arab, A., Hajisalimi, Z., Ahmadi, S. J., Bakhtiari, A. (2017): Developing the Persian Gulf Wave Forecasting System, Journal of Coastal and arine Engineering, Vol. 1, Issue 1.
[38]   S. A. Haghshenas; Ghader, S.; Yazgi, D.; Delkhosh, E.; Rashedi Birgani.; Razavi Arab, A.; N., Hajisalimi, Z., Nemati, M. H.; Soltanpour, M., and Jedari Attari, M., (2018). Iranian Seas Waters Forecast - Part I: An Improved Model for The Persian Gulf. . In: Shim, J.-S.; Chun, I., and Lim, H.S. (eds.), Proceedings from the International Coastal Symposium (ICS) 2018 (Busan, Republic of Korea). Journal of Coastal Research, Special Issue No. 85, pp. 1216-1220. Coconut Creek (Florida), ISSN 0749-0208.